\documentclass[fleqn,12pt,twoside]{article}
\usepackage{espcrc1}
\usepackage{graphicx}

\usepackage{latexsym}
\usepackage{epsfig}
\usepackage{bm}
\usepackage{amssymb,amsmath}

\begin{document}

\title{\boldmath Indication of an excited hyperon state in \textit{pp} collisions with ANKE at COSY-J\"ulich}
\author{
  I.~Zychor\address{The Andrzej So{\l}tan Institute for Nuclear Studies, 05-400 \'Swierk, Poland},
  V.~Koptev\address[GATCHINA]{Petersburg Nuclear Physics Institute, 188350 Gatchina, Russia},
  M.~B\"uscher\address[FZJ]{Institut f\"ur Kernphysik, Forschungszentrum J\"ulich, 52425 J\"ulich, Germany},
  A.~Dzyuba\addressmark[FZJ],
  I.~Keshelashvili\addressmark[FZJ],
  V.~Kleber\address{Institut f\"ur Kernphysik, Universit\"at zu K\"oln, 550937 K\"oln, Germany},
  R.~Koch\addressmark[FZJ],
  S.~Krewald\addressmark[FZJ],
  Y.~Maeda\addressmark[FZJ],
  S.~Mikirtichyants\addressmark[GATCHINA],
  M.~Nekipelov\addressmark[GATCHINA]\addressmark[FZJ],
  H.~Str\"oher\addressmark[FZJ] 
  and C.~Wilkin\address{University College London, London WC1E 6BT, U.K.}
}
\date{\today}

\maketitle

\begin{abstract}
The reaction $pp \rightarrow pK^+ Y$ has been studied with the ANKE
spectrometer at COSY-J\"ulich in order to investigate heavy hyperon
production.  The missing mass spectra $MM(pK^+)$ have been analyzed
and compared with Monte Carlo simulations.  Indications for a hyperon
resonance $Y^{0*}(1480)$ have been found.
\end{abstract}

\section{Introduction}

The production and properties of hyperons have been studied for more
than 50 years, mostly in pion and kaon induced reactions.
Hyperon production in pp collisions has been investigated close to
threshold at SATURNE (Saclay, France) and COSY-J\"ulich.
Reasonably
complete information on $\Lambda(1116)$, $\Sigma^0(1192)$,
$\Sigma^0(1385)$, $\Lambda(1405)$ and $\Lambda(1520)$ can be found in
the literature \cite{PDG} although for the $\Lambda(1405)$, in spite
of rather high statistics achieved (the total world statistics is
several thousand events), there are still open questions concerning
the nature of this resonance: is it a singlet \textit{qqq} state in
the frame of SU(3) or a quark-gluon \textit{(uds-q)} hybrid, or a KN
bound state? The $\Sigma(1480)$ hyperon is not well established yet.
In the 2004 Review of Particle Physics it is described as a 'bump'
with unknown quantum numbers.

We have investigated whether additional information on hyperon production 
might be obtained from proton-proton interactions at low energies. 

\section{Experiment and simulations}

The experiment has been performed at the Cooler Synchrotron COSY at
the Research Center J\"ulich (Germany)~\cite{COSY}.  COSY is a medium
energy accelerator and storage ring for both polarized and unpolarized
protons and deuterons.  The measurements were performed at a proton
beam momentum of 3.65~GeV/c incident on a a hydrogen cluster--jet
target.  The average luminosity during the measurements was
$L=(1.38\pm 0.15)\times 10^{31}\,\textrm{s}^{-1}\,\textrm{cm}^{-2}$.
\begin{figure}[ht]
\centerline{\psfig{file=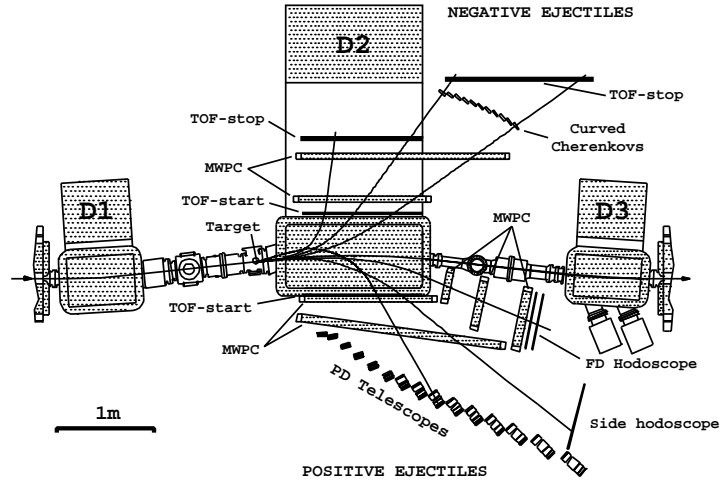,width=9.5cm}}
\vspace*{-0.9cm}
  \caption{ANKE spectrometer and detectors.}
  \label{fig:ANKE}
\end{figure}
\vspace*{-0.7cm}
The ANKE magnetic spectrometer~\cite{ANKE_NIM} used in the experiment
consists of three dipole magnets. The central C--shaped spectrometer D2,
placed downstream of the target, separates the reaction products from
the circulating COSY beam.  The ANKE detection system, comprising
range telescopes, scintillation counters and multi--wire proportional
chambers, simultaneously registers both positively and negatively
charged particles and measures their momenta~\cite{K_NIM}.

\begin{figure}[hb]
\vspace*{-0.7cm}
\centerline{\psfig{file=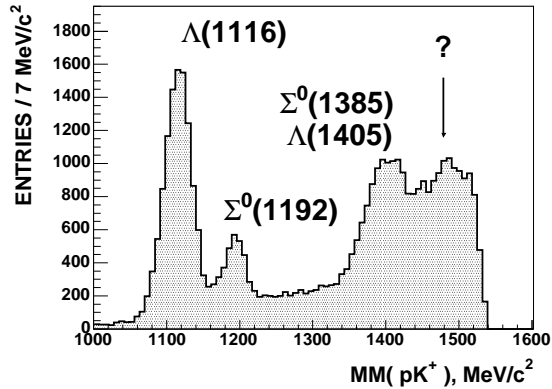,width=9.0cm}}
\vspace*{-1.2cm}
  \caption{
   Missing mass $MM(pK^+)$ distribution measured in the reaction 
   3.65~GeV/c~\mbox{$pp \rightarrow pK^+Y$} with indicated neutral hyperons.}
  \label{fig:only_pK}
\end{figure}

\begin{figure*}[Ht]
\psfig{file=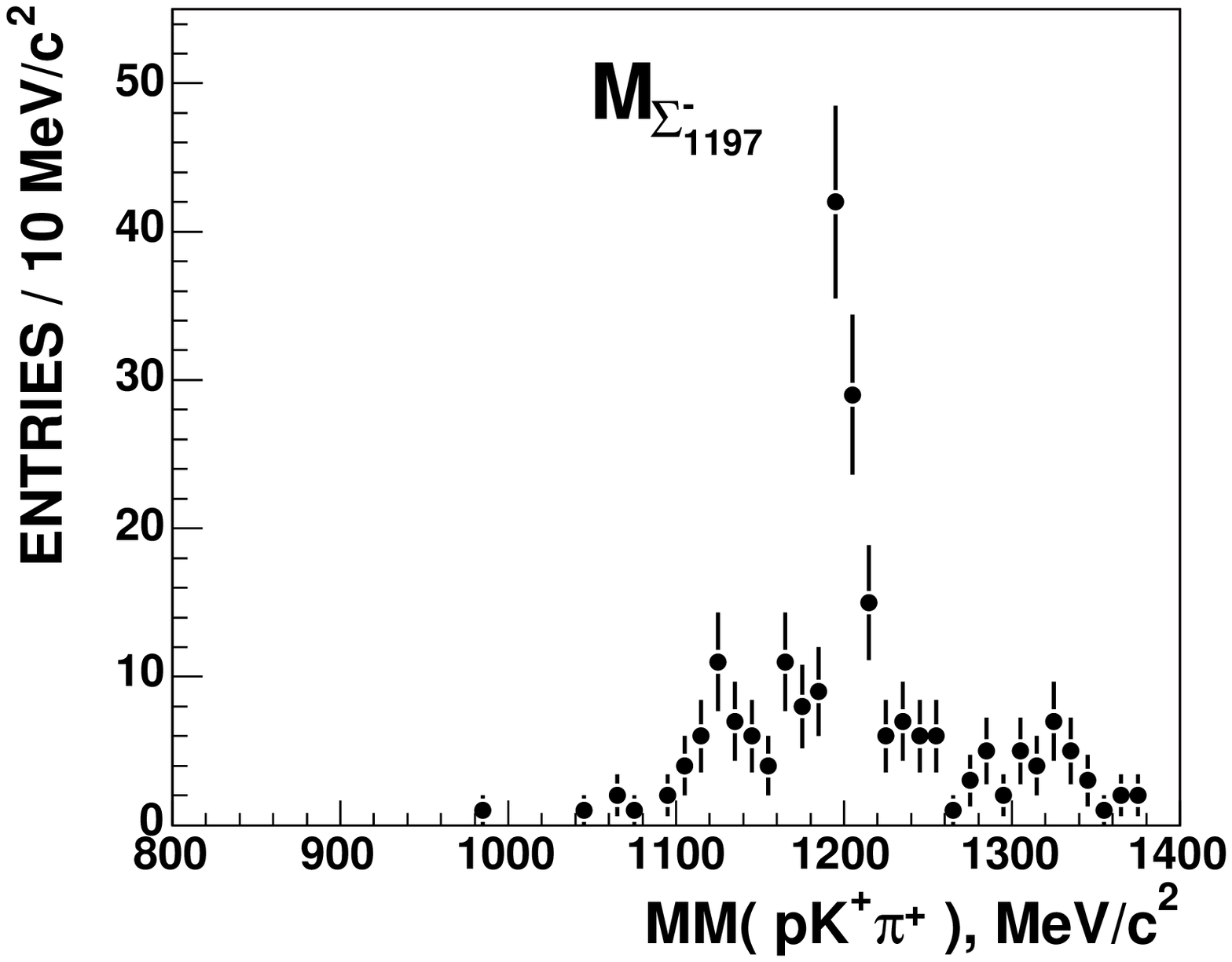,width=7.9cm}
\hspace*{-1.0cm}
\psfig{file=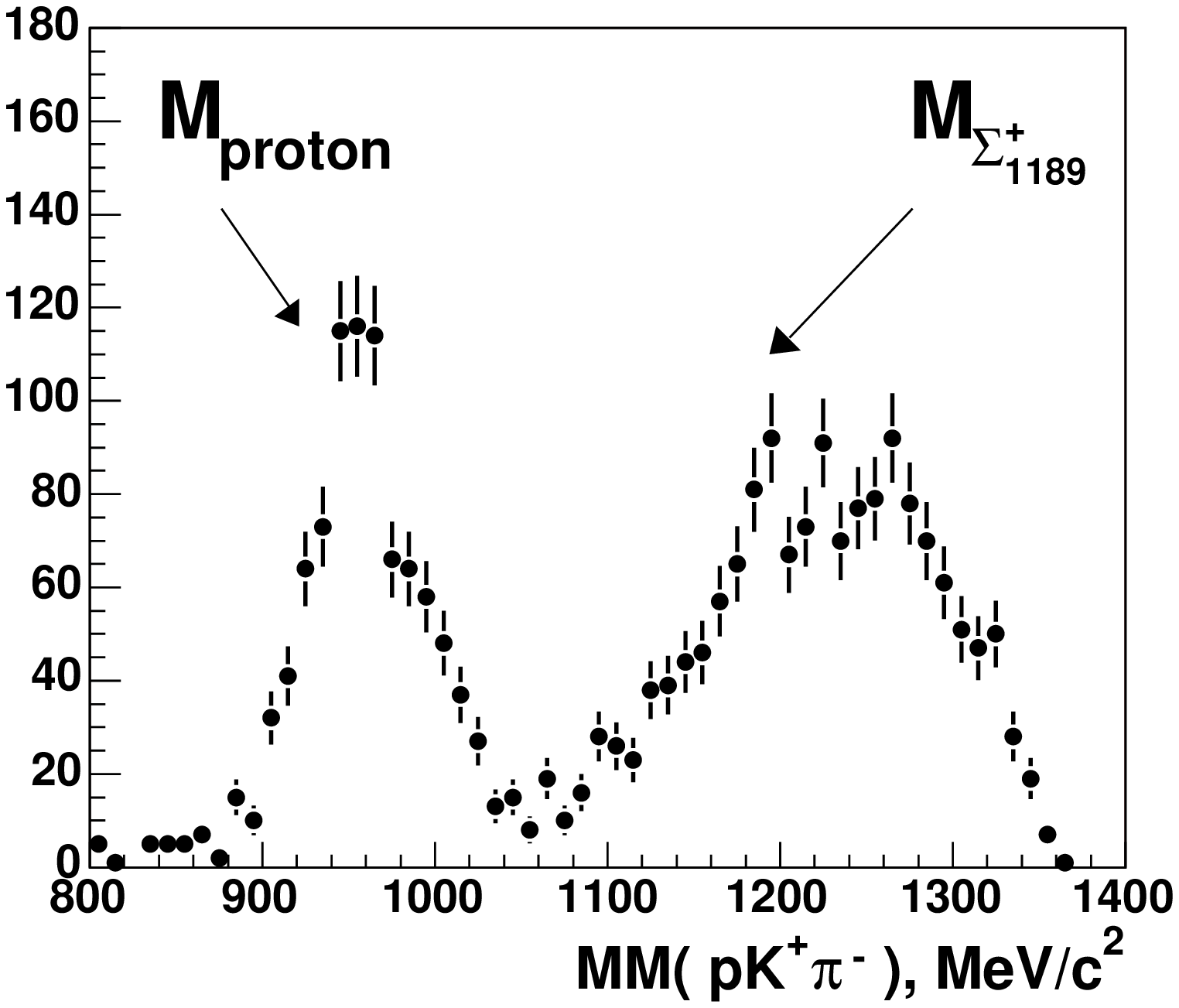,width=7.9cm}
\vspace*{-1.5cm}
  \caption{Missing mass spectra $MM(p{K}^+\pi)$ for the reaction \mbox{$pp \rightarrow
    pK^+\pi^+X^-$}(left) and \mbox{$pp \rightarrow pK^+\pi^-X^+$}(right).} 
  \label{fig:MM3}
\end{figure*}

At a COSY beam momentum of 3.65~GeV/c hyperons $Y$ with masses up to
$\sim$1540 MeV/c$^2$ can be produced in the reaction $pp \rightarrow
pK^+ Y$.  
In Fig.~\ref{fig:only_pK} the missing
mass $MM(pK^+)$ spectrum is shown for the case that a $K^+$~meson in
coincidence with a proton is detected with ANKE. While the
ground-state hyperons are clearly seen in the spectrum, the heavier
ones are on top of a broad background which can mainly be attributed
to the production of additional pions. Thus the unambigous
identification of these hyperons requires the detection of additional
coincident particles from their decays.
A final state comprising a proton, a positively charged kaon, a pion
of either charge and an unidentified residue X was investigated in the
reaction $pp \rightarrow p{K}^+Y \rightarrow p{K}^+\pi^\pm X^\mp$.  A
missing mass $MM(p{K}^+\pi^+)$ spectrum in the reaction $pp\to p{K}^+\pi^+X^-$ shows a
flat background with a peak at approximately 1195~MeV/$c^2$ (left part
in Fig.~\ref{fig:MM3}).
The peak corresponds to the decay $Y\to\pi^+\Sigma^-(1197)$.  In the
charge--mirrored \mbox{$pp \rightarrow pK^+\pi^-X^+$} case, the
$\pi^-$ may originate from different sources, \textit{e.g.}\ a decay
with the $\Sigma^+(1189)$ or a secondary decay of $\Lambda \to
p\pi^-$, arising from the major background reaction $pp \rightarrow
pK^+\Lambda \rightarrow pK^+\pi^-p$. Protons from this reaction can be
easily rejected by cutting the missing mass $MM(pK^+\pi^-)$ around the proton mass
(right part in Fig.~\ref{fig:MM3}). Nevertheless the missing mass
distribution for the $(\pi^-X^+)$-final state is more complicated.
\begin{figure*}[Hb]
\psfig{file=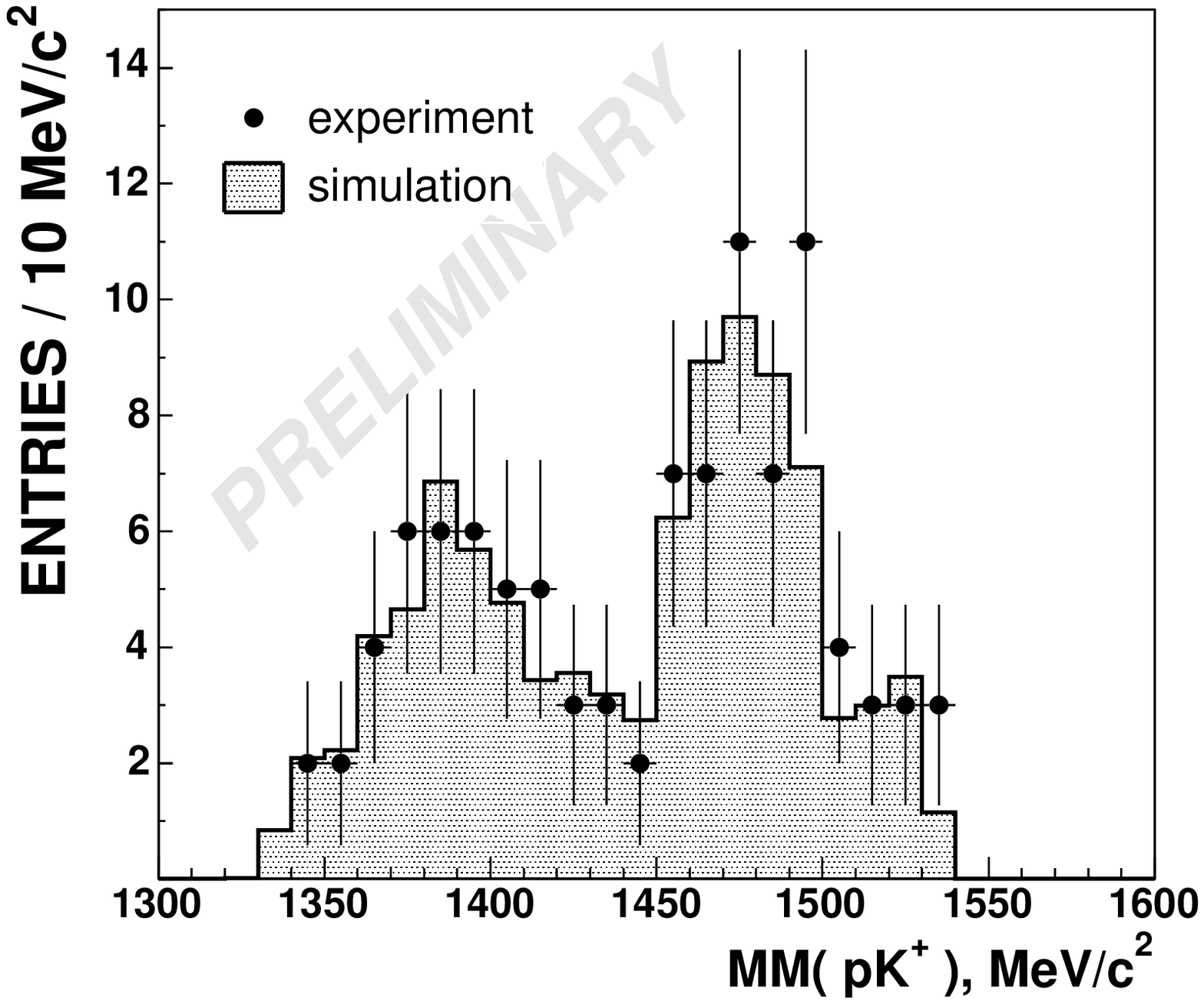,width=7.8cm}
\hspace*{-1.0cm}
\psfig{file=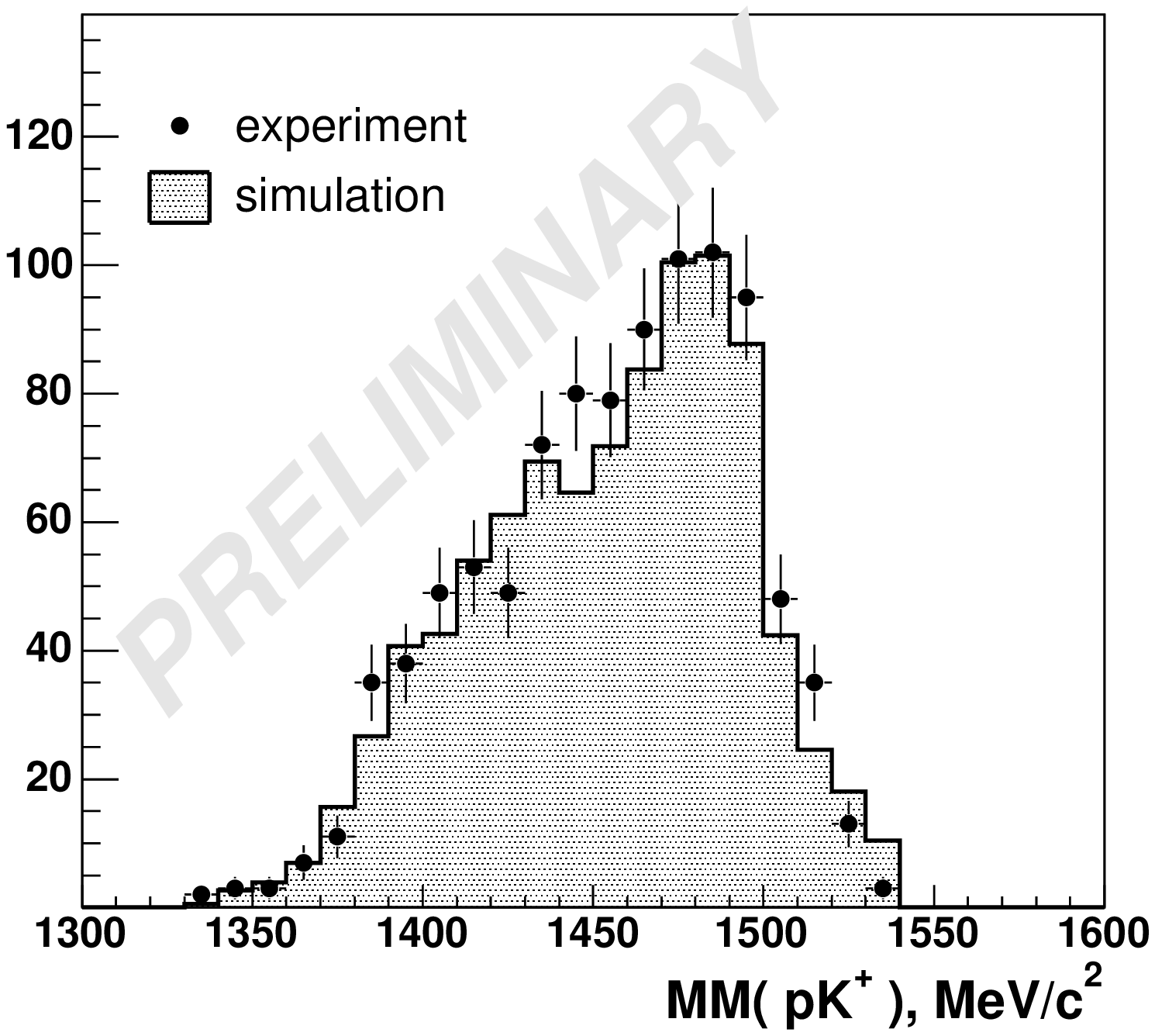,width=7.8cm}
\vspace*{-1cm}
  \caption{ Experimental and simulated missing mass $MM(pK^+)$ spectra
  for the reaction $pp \rightarrow pK^+\pi^+X^-$(left) and $pp
  \rightarrow pK^+\pi^-X^+$(right). The shaded histogram shows the
  fitted Monte Carlo simulations.  } \label{fig:MM2}
\end{figure*}

If only events around the $\Sigma$ mass are selected, then the missing
mass spectrum $MM(pK^+)$ in the reaction $pp\to p {K}^+ \pi^+ X^-$
shows two peaks, see left part in Fig.~\ref{fig:MM2}. One of them
corresponds to the contribution of $\Sigma^0(1385)$ and
$\Lambda(1405)$ hyperons. The second peak is located at a mass
$\sim$1480~MeV/c$^2$. In the $\pi^-X^+$ case, the distribution also
peaks at 1480~MeV/c$^2$, see right part in Fig.~\ref{fig:MM2}.

We have assumed that the measured missing mass $MM(pK^+)$ spectra can
be explained by the production of hyperon resonances and non-resonant
contributions.  Detailed Monte Carlo simulations have been performed
including the production of well established excited hyperons
($\Sigma^0(1385), \Lambda(1405)$ and $\Lambda (1520)$) and
non-resonant contributions like $pp \rightarrow pK^+\pi X$ and $pp
\rightarrow pK^+\pi\pi X$; $X$ denotes any hyperon which could be
produced in the experiment.  From the comparison of measured and
simulated missing mass distributions it turned out that it is
necessary to include another excited hyperon $Y^{0*}$ with a mass
$M(Y^{0*})= (1480\pm15)~\textrm{MeV/c}^2$ and a width $\Gamma(Y^{0*})
= (60\pm 15)~\textrm{MeV/c}^2$.

\section{Conclusions}
We have found an indication of a neutral hyperon resonance $Y^{0*}$ 
produced in proton--proton
collisions at 3.65~GeV/c and
decaying into $\pi^+ X^-$ and $\pi^- X^+$ final states. Its
parameters are $M(Y^{0*}) = (1480\pm 15)~\textrm{MeV/c}^2$ and
$\Gamma(Y^{0*}) = (60\pm 15)~\textrm{MeV/c}^2$ though, since it is
neutral, it can be either a $\Lambda$ or $\Sigma$~hyperon.
The statistical significance of a
signal is at the level of 4.8 standard deviations.
The production cross section is of the order of few hundred nanobarns.
It seems to be difficult to
integrate the low mass $Y^{0*}$ hyperon within the existing
classification of 3q-baryons \cite{EPJA10_447,CapIs}.
On the basis of available data we cannot decide whether it is a
3--quark baryon or an exotic state, although some preference
towards its exotic nature may be deduced from theoretical
considerations 
\cite{NPB145_119,Azimov,Arndt,Jaffe_Wilczek_PRD69_114017,PRD69_094009_antidecuplet,PLB582_49_quark_mass}.

Further studies are required to determine its quantum numbers. At
ANKE, using a deuterium cluster target and spectator proton
tagging, one can search for the charged $Y^{-*}$ hyperon in the
reaction $pn \rightarrow pK^+ Y^{-*} \rightarrow pK^+ \pi^- X^0$.
The investigation of $Y^*$ decays with photons in the final state
is foreseen with the WASA detector at COSY~\cite{WASA}.\ \\

This work has been supported by: FFE Grant (COSY-78, nr 41553602),
BMBF (WTZ-RUS-211-00, 691-01, WTZ-POL-015-01, 041-01),
DFG (436 RUS 113/337, 444, 561, 768),
Russian Academy of Sciences (02-04-034, 02-04034, 02-18179a, 02-06518).

\end{document}